\documentclass[twocolumn,showpacs,preprintnumbers,amsmath,amssymb,floatfix]{revtex4}

\usepackage{epsfig,psfrag}
\usepackage{dcolumn}
\usepackage{bm}
\usepackage{graphicx}
\usepackage{color}

\begin{document}

\title{Anomalous Josephson current through a spin-orbit coupled quantum dot}
\author{A. Zazunov,$^1$ R. Egger,$^1$ T. Jonckheere,$^2$ and T. Martin$^{2,3}$}
\affiliation{$^1$~Institut f\"ur 
Theoretische Physik, Heinrich-Heine-Universit\"at,
D-40225  D\"usseldorf, Germany\\ $^2$~Centre de Physique 
Th{\'e}orique, Campus de Luminy, case 907, F-13288 Marseille, France\\
$^3$Universit{\'e} de la M{\'e}dit{\'e}rann{\'e}e,  F-13288 Marseille, France}
\date{\today}

\begin{abstract}
For a general model of a mesoscopic multi-level quantum dot, we determine the
necessary conditions for the existence of an anomalous Josephson 
current with spontaneously broken time-reversal symmetry.  They 
correspond to a finite spin-orbit coupling, a
suitably oriented Zeeman field, and the dot being a chiral conductor. 
We provide analytical expressions for the anomalous supercurrent
covering a wide parameter regime. 
\end{abstract}

\maketitle

\textit{Introduction.---}
The Josephson effect, where an equilibrium supercurrent 
flows through a junction between two superconductors held at
phase difference $\phi$, is of fundamental importance in 
condensed matter physics, quantum information science, 
microelectronic applications, and metrology \cite{jj}.
It has recently attracted renewed interest in mesoscopic and nanoscale
junctions after the experimental demonstration
of gate-tunable Josephson currents through junctions with just a few
relevant electronic levels (``quantum dot'') in a variety of material 
systems, e.g., InAs nanowires \cite{doh}, 
the 2D electron gas in semiconductors \cite{takayanagi},
and carbon nanotubes \cite{eichler}.
One important novel aspect arises 
because the spin-orbit interaction (SOI) strength $\alpha$ due 
to structural and bulk inversion asymmetries is often significant 
\cite{winkler,nitta}, and theoretical work has therefore started to 
address SOI effects on the Josephson current in such devices 
\cite{bezuglyi,krive1,feigelman,luca,beenakker,feinberg,buzdin}.
So far this activity has mainly focused on 0- and $\pi$-junctions
(positive or negative critical current $I_c$, respectively).
A remarkable prediction concerns the possibility for an 
\textit{anomalous supercurrent}\ $I_a$, flowing even at zero phase 
difference ($\phi=0$) if both a (suitably oriented) Zeeman field ${\bm b}$
and the SOI are present \cite{krive1,feinberg,buzdin}.
For $\phi=0$, the Hamiltonian is invariant under time-reversal symmetry (TRS) 
even when $\alpha\ne 0$ and ${\bm b}\ne 0$, and the anomalous supercurrent 
thus spontaneously breaks TRS, which otherwise 
enforces $I(\phi)=-I(-\phi)$ and hence $I_a=0$ \cite{jj}.
Anomalous supercurrents were first predicted in 
unconventional superconductors \cite{larkin},
but were never observed there.  In the tunneling limit, 
where the conventional current-phase relation (CPR) 
reads $I(\phi)=I_c\sin\phi$, this
is equivalent to a \textit{phase shift}\ $\phi_0$, i.e.,
$I(\phi)=I_c\sin(\phi+\phi_0)$ and thus $I_a=I_c\sin\phi_0$.  
The phase shift in such a ``$\phi_0$-junction'' could 
be observed in a SQUID containing one 0- and 
one $\phi_0$-junction via the shift of the diffraction pattern, or
as spontaneous current in a superconducting ring containing a 
$\phi_0$-junction.  Both effects are tunable by external gate 
voltages (affecting the SOI), Zeeman fields, and by an orbital magnetic flux.
Moreover, a $\phi_0$-junction can also act as a 
superconducting rectifier \cite{feinberg}.  

Mesoscopic systems contacted by conventional $s$-wave BCS
superconductors could then yield a new class of systems
with spontaneously broken TRS, and therefore exhibit
anomalous supercurrents.  Recent works have started to address 
this point. First, Ref.~\cite{krive1} considered a long ballistic
one-dimensional Rashba quantum wire, where $I_a\ne 0$ is
tied to the Zeeman effect and to the difference between the 
velocities of right- and left-moving electrons. 
However, for physically realizable $\alpha$, the reported $I_a$ values
turn out to be extremely small, $I_a\propto \alpha^4$,
or are most likely inaccessible in experiments. 
In a mainly numerical study \cite{feinberg},  
the anomalous Josephson effect was also found for a multi-channel
spin-polarizing quantum point contact. 
Finally, Buzdin \cite{buzdin} reported 
$I_a\neq 0$ in junctions containing a noncentrosymmetric ferromagnet
as a weak link. 
There is clearly a need to systematically classify all ingredients 
necessary to observe the anomalous Josephson effect.  In this work, 
we compute the Josephson current through a generic phase-coherent 
mesoscopic system (an arbitrary multilevel quantum dot).
Analytical predictions for $I_a$ are provided for a physically 
important parameter regime. In addition, these results are supported
by numerics. A necessary condition for $I_a\ne 0$ emerges from our
study: the quantum dot must be a \textit{chiral conductor}, 
see Eq.~(\ref{condition}) below.
This requirement was implicit in Ref.~\cite{krive1} but is
apparently violated \cite{footnote3} in Ref.~\cite{buzdin}.

\textit{Model and exact solution.---} 
As a generic model for the Josephson current through a quantum dot, 
we consider $H=H_L+H_R+H_T+H_D$, with two identical $s$-wave BCS 
superconductors ($H_{j=L/R}$) of gap $\Delta$ held at phase
 difference $\phi$ and tunnel-coupled ($H_T$) to the 
quantum dot.  With lead fermion operators $c_{L/R,{\bm k}\sigma}$ 
for spin $\sigma=\uparrow,\downarrow$ and momentum ${\bm k}$, 
the BCS Hamiltonian reads (we often put $\hbar=1$ and, for simplicity, 
take the zero-temperature limit) 
\[
H_j = \sum_{{\bm k}\sigma} \frac{k^2}{2m} c^\dagger_{j\bm k\sigma} 
c^{}_{j\bm k\sigma} + \sum_{\bm k} \left( \Delta e^{\mp i \phi/2}
c^\dagger_{j{\bm k} \uparrow} c^\dagger_{j(-{\bm k})\downarrow}
+ {\rm h.c.}\right).
\]
For $\alpha=0$ and ${\bm b}=0$, the closed dot can be described in terms
of real-valued eigenfunctions $\chi_n({\bm r})$ [with ${\bm r}=(x,y,z)$]
for eigenenergy $\epsilon_n$, where $n=1,\ldots,N$ labels the relevant 
dot orbitals.  Using fermion operators $d_{n\sigma}$ for these orbitals
and disregarding strong correlation effects,
the dot Hamiltonian in the presence of SOI and Zeeman field can be 
written in the form (the Pauli matrices $\sigma_{x,y,z}$ act in spin space, 
spin indices are kept implicit, and we absorb the magnetic
factor $g\mu_B/2$ into ${\bm b}$)
\begin{equation}
H_D = \sum_{n} d_{n}^\dagger \ [ \epsilon_n + 
 {\bm b} \cdot {\bm\sigma} ] \ d_n^{}
-i \sum_{nn'} d_n^\dagger \ {\bm a}_{nn'}  \cdot {\bm \sigma}  \ d_{n'}^{}.
\end{equation} 
We consider both the Rashba and Dresselhaus SOI due to 
structural and bulk inversion asymmetry, respectively, with the electric field
pointing in $z$ direction. The SOI is encoded in the
antisymmetric matrices ${\bm a}_{x,y}$ forming the vector \cite{luca}
\begin{equation}\label{adef}
{\bm a}_{nn'}= \frac{\alpha}{m}  \int d{\bm r} \ \chi_n({\bm r})
\left( \begin{array}{c} \sin(\theta) \partial_x + \cos(\theta)\partial_y\\
-\cos(\theta) \partial_x -\sin(\theta)\partial_y \end{array}\right)
\chi_{n'}({\bm r}) .
\end{equation}
The overall SOI strength is represented by the characteristic inverse length
$\alpha$, while the angle $\theta=0$ 
corresponds to a pure Rashba and $\theta=\pi/2$
to a pure Dresselhaus case \cite{winkler}. 
The tunneling Hamiltonian is 
\begin{equation}\label{htun}
H_T = \sum_{j=L/R, {\bm k} n \sigma} t^{}_{jn} c^\dagger_{j{\bm k}\sigma} 
d^{}_{n\sigma} + {\rm h.c.},
\end{equation}
where we make the inessential assumption of ${\bm k}$-independent tunneling
matrix elements.  It is useful to define the 
Hermitian $N\times N$ hybridization matrices (in dot level space), 
\begin{equation}\label{glr}
{\bm \Gamma}_{j=L/R,nn'} = \pi \nu \ t^*_{jn} t^{}_{jn'},
\end{equation}
where $\nu$ is the normal-state density of states in the
leads.  We find that the anomalous supercurrent is most pronounced if the
Zeeman field points along a direction defined by the SO angle $\theta$, 
namely ${\bm b}= B ( \cos\theta, - \sin\theta, 0 )^T$.
When the magnetic field is orthogonal to this, 
the anomalous supercurrent is absent.
The above model ignores orbital magnetic fields.  For the important case
of a 2D dot, this is justified by choosing an in-plane magnetic field
as above.

Integrating out all fermionic degrees of freedom, the Josephson current for
arbitrary system parameters is obtained:
\begin{equation}\label{current}
I(\phi) = -\frac{2e}{h} \int_0^\infty d\omega \ \partial_\phi \ {\rm tr} \ln
{\bm S}(\omega)
\end{equation}
with a $4N\times 4N$ matrix ${\bm S}(\omega)$.  Using auxiliary Pauli 
matrices $\tau_{x,z}$, the matrix ${\bm S}$ can be expressed as \cite{luca}
\begin{eqnarray*}
{\bm S}&=& -i\omega \left( 1+ \frac{{\bm \Gamma}_L+{\bm \Gamma}_R}
{\sqrt{\omega^2+\Delta^2}} \right) + {\bm E}\sigma_z\tau_z  + {\bm Z}
+\frac{\Delta}{\sqrt{\omega^2+\Delta^2}} \\ &\times& 
\left[ ({\bm \Gamma}_L+{\bm\Gamma}_R) \cos(\phi/2) \sigma_x\tau_z
+ ({\bm  \Gamma}_L-{\bm\Gamma}_R) \sin(\phi/2) \sigma_y \right],
\end{eqnarray*}
where ${\bm E} \equiv {\rm diag}( \tilde\epsilon_1, \ldots,\tilde\epsilon_N)$
with $\tilde\epsilon_n=\epsilon_n-\alpha^2/2m$ contains the dot level energies.
The Zeeman field and the SOI are encoded in the 
$\omega$-independent $4N\times 4N$ matrices 
\begin{eqnarray}\nonumber
{\bm Z} &=& (i {\bm A}_x \sigma_x + {\bm B}_y \sigma_y)\tau_x +
(i {\bm A}_y \sigma_x - {\bm B}_x \sigma_y)\tau_y \\ \label{zdef}
&+& {\bm B}_z \tau_z + i{\bm A}_z \sigma_z ,
\end{eqnarray}
with the vector of real anti-symmetric matrices \cite{footgauge}
\begin{equation}\label{Adef}
{\bm A}_{nn'} = \frac{\alpha}{m} \int d{\bm r} \
\chi_n \partial_y \chi_{n'}
\left( \begin{array}{c}- \cos\theta [1-2\cos(2\theta)\sin^2(\alpha x)]\\
\sin\theta [1+2\cos(2\theta)\sin^2(\alpha x)]\\
\cos(2\theta)\sin(2\alpha x) \end{array}\right).
\end{equation}
The matrices ${\bm B}_{x,y,z}$ are real and symmetric, and 
for the above Zeeman field given by
\begin{eqnarray}\label{Bdef}
{\bm B}_{nn'} =  B \int d{\bm r} \ \chi_n \chi_{n'} 
\hspace{4cm}\nonumber \\
\times \, \left( \begin{array}{c}
\cos\theta\cos^2(\alpha x)-\cos(3\theta)\sin^2(\alpha x)\\ 
-\sin\theta\cos^2(\alpha x)-\sin(3\theta)\sin^2(\alpha x)\\
-\cos(2\theta)\sin(2\alpha x) \end{array}\right).
\end{eqnarray}
One can verify from the above expressions that no 
anomalous supercurrent can exist in the absence of either the SOI or 
the Zeeman field.

\textit{Analytical approach.---}
We now derive the anomalous supercurrent $I_a$ in
the most relevant limit of weak SOI and weak Zeeman field, where $I_a
\propto \alpha B$ is small. Moreover, the derivation below assumes 
that the off-diagonal entries in the ${\bm \Gamma}_{L/R}$ matrices
(\ref{glr}) are small against the diagonal entries.
This condition is met in most cases of practical interest, e.g.,
if one dot level is resonantly coupled to the leads and all
other levels are only weakly coupled, or when quasi-random phase shifts
between different $t_{jn}$ have to be taken into account.
We then consider the limit $\phi\to 0$, where 
under the above conditions, it makes sense to write 
${\bm S}={\bm S}_0 + {\bm S}_1$ in Eq.~(\ref{current}).
The ``leading'' part is diagonal in dot level space,
\begin{equation} \label{s0}
{\bm S}_0 = -i\omega + {\bm E}\sigma_z\tau_z 
 + \frac{-i\omega+\Delta\sigma_x\tau_z}{\sqrt{\omega^2+\Delta^2}} \
{\bm \Gamma}_0 ,
\end{equation}
with ${\bm \Gamma}_0= {\rm diag}({\bm \Gamma}_L+{\bm\Gamma}_R)$. 
Similarly, using ${\bm \Gamma}_1={\bm \Gamma}_L+
{\bm\Gamma}_R-{\bm \Gamma}_0$, the ``perturbation'' part is
\begin{equation}\label{s1}
{\bm S}_1 = {\bm Z} + \frac{(\phi/2) \Delta 
({\bm\Gamma}_L-{\bm\Gamma}_R) \sigma_y + [-i\omega+\Delta 
\sigma_x\tau_z] {\bm \Gamma}_1} {\sqrt{\omega^2+\Delta^2}} 
\end{equation}
with ${\bm Z}$ given in Eq.~(\ref{zdef}).
The anomalous supercurrent then follows
by expansion of the tracelog in Eq.~(\ref{current}),
\begin{equation} \label{expand}
I_a = \frac{2e}{h}\sum_{n=1}^\infty \frac{(-1)^n}{n}
\int_0^\infty d\omega \ 
 \partial_\phi \ 
{\rm tr}( {\bm S}_0^{-1}
 {\bm S}_1 )^n .
\end{equation}
Using Eq.~(\ref{zdef}), straightforward but lengthy algebra shows
that both the $n=1$ and $n=2$ contributions always vanish. 
 The leading contribution to $I_a$ then comes from $n=3$, where the part 
$\propto {\bm \Gamma}_1$ in ${\bm S}_1$, see Eq.~(\ref{s1}),
does not contribute at all.   
In the end, we arrive at the analytical expression 
\begin{eqnarray}\nonumber 
I_a &=& \frac{8e \Delta^2}{h} \int_0^\infty 
\frac{d\omega}{\omega^2+\Delta^2} \ {\rm tr}_d \Biggl( 
[ {\bm \Gamma}_R,{\bm \Gamma}_L ]_- {\bm D}^{-1} ({\bm b}\cdot{\bm A})
\\  \label{final} &\times& {\bm D}^{-1} \left [ 1-
4\omega^2 \left(1+\frac{{\bm\Gamma}_0}{\sqrt{\omega^2+\Delta^2}}\right)
{\bm D}^{-1} \right ] \Biggr)
\end{eqnarray}
with the diagonal matrix
${\bm D}= \omega^2 + {\bm E}^2 + {\bm \Gamma}_0^2 + 
\frac{2\omega^2}{\sqrt{\omega^2+\Delta^2}} {\bm \Gamma}_0.$
The trace operation tr$_d$ extends over dot level space only.  
Since ${\bm b}\cdot{\bm A}\propto \alpha B$, we see that 
$I_a\propto \alpha B$ as expected.  Apart from having a finite SOI and an
appropriately oriented Zeeman field, an additional condition must be 
satisfied in order to have $I_a\ne 0$:
\begin{equation}\label{condition}
[ {\bm \Gamma}_R,{\bm \Gamma}_L ]_- \ne 0,
\end{equation}
which implies chirality for transport through the quantum dot. 
{}From numerical studies of the full current (\ref{current}), we find
that Eq.~(\ref{condition}) is a necessary condition for $I_a\ne 0$
for arbitrary other parameters, and hence is not restricted by the
conditions under which Eq.~(\ref{final}) has been obtained. 
As a consequence, a single-level dot ($N=1$) can never allow for an
anomalous supercurrent, and at least two relevant orbitals are required.
Finally, we remark that Eq.~(\ref{final}) can be further simplified for
${\rm max}({\bm \Gamma}_{nn}) \gg \Delta$, where we obtain
\begin{equation}\label{phi0}
I_a = \frac{2e\Delta}{\hbar} {\rm tr}_d \left(
({\bm\Gamma}^2_0+{\bm E}^2)^{-1} [ {\bm \Gamma}_R,{\bm \Gamma}_L ]_- 
({\bm\Gamma}^2_0+{\bm E}^2)^{-1} {\bm b}\cdot{\bm A} \right).
\end{equation}

\begin{figure}
\scalebox{0.32}{\includegraphics{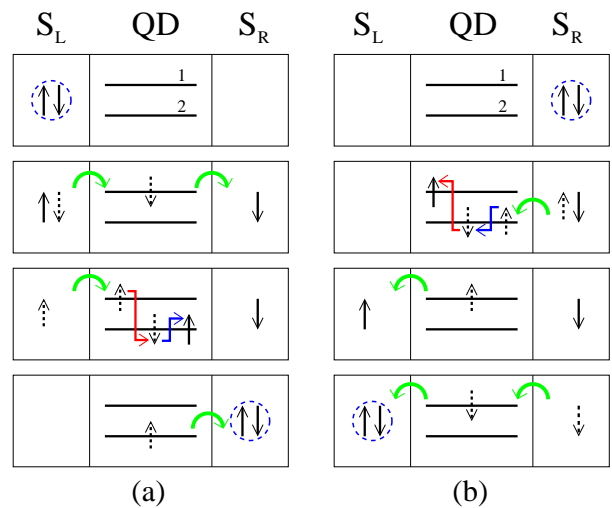}}
\caption{ \label{f1} (Color online)
Schematic picture of transfer of a Cooper pair through a two-level dot.
(a) Contribution to $t_{L\to R}$ yielding an 
anomalous supercurrent. (b) Reverse process contributing to
$t_{R\to L}$. Top and bottom panels represent initial and final 
states, respectively,
which are connected by a sequence of the intermediate virtual states.
Solid arrows indicate transitions due to tunneling (green, connecting leads
and dot), spin-orbit (red, connecting different dot levels) and 
Zeeman (blue) coupling. For details see text.
}
\end{figure}

\textit{Basic explanation.---}
Having established that spontaneously broken TRS is indeed
possible in this system, see Eq.~(\ref{final}), we now give
an intuitive physical argument to understand the origin of this effect.  
Consider the transfer of a Cooper pair through a quantum dot with $N=2$ 
levels for $\phi=0$, schematically shown in Fig.~\ref{f1}. 
The Cooper pair has amplitude $t_{L\to R}$ 
($t_{R\to L}$) for transfer from left to right (right to left). 
For simplicity, we put $\theta=0$ and assume real tunneling
amplitudes $t_{jn}$ in Eq.~(\ref{htun}).  To lowest order in $\alpha B$, 
SOI and Zeeman field combine to the term
$H' = (i{\bm A}_x+{\bm B}_x) \sigma_x$, with $2\times 2$ matrices
in dot level space,
${\bm A}_x = A \left(\begin{array}{cc} 0 & 1 \\ -1 & 0 \end{array}\right)$,
where $A\propto \alpha$, 
and ${\bm B}_x = B \left(\begin{array}{cc} 1 & 0 \\ 0 & 1 \end{array}\right)$.
To lowest order in the $t_{jn}$ and in $H'$,
processes like the one sketched in Fig.~\ref{f1}(a) generate a 
contribution to $t_{L\to R}$. Specifically, the process  in Fig.~\ref{f1}(a)
yields
\begin{equation}\label{special}
\delta t_{L\to R} = (t_{L1} t_{R1}) \ (t_{L1} \ i A B \ t_{R2}),
\end{equation}
where the $\downarrow$-electron causes the first factor while the
$\uparrow$-electron switches levels via the product of ${\bm A}_x$
and ${\bm B}_x$ processes.  In effect, the SOI and the Zeeman field 
conspire to produce an \textit{effective orbital magnetic field}\
via such processes, which then explains the broken TRS.
With the group velocity $v$, the anomalous supercurrent contribution 
from Eq.~(\ref{special}) is $\delta I_a  \propto v A B \ {\bm\Gamma}_{L,11}
{\bm\Gamma}_{R,12}$.  
Consider next the reverse process, see Fig.~\ref{f1}(b), which yields
$\delta t_{R\to L}= (t_{R2} B (-iA) t_{L1})
\ (t_{R1} t_{L1})$, and therefore causes the 
\textit{same} current contribution
$\delta I_a \propto (-v)\cdot (-A) B {\bm\Gamma}_{L,11} {\bm\Gamma}_{R,12}$.
Summing up all relevant processes, we find 
\begin{eqnarray} \label{finalan}
I_a &\propto & B(t_{L1} t_{R1}+t_{L2}t_{R2}) (t_{L1} A t_{R2}
+ t_{L2}(-A) t_{R1}) \\ \nonumber
&=& AB \left[ ({\bm\Gamma}_{L,11}-{\bm\Gamma}_{L,22}) {\bm\Gamma}_{R,12} 
-({\bm\Gamma}_{R,11}-{\bm\Gamma}_{R,22}) {\bm\Gamma}_{L,12} \right].
\end{eqnarray}
Note that $I_a\ne 0$ precisely when Eq.~(\ref{condition}) holds,
as follows by explicitly computing the commutator in Eq.~(\ref{condition})
for $N=2$.  The above arguments resemble the justification for 
$\pi$-junction behavior in 
quantum dots with Coulomb blockade \cite{spivak}, and can in fact
be applied to show that $I_a\ne 0$ under the specified conditions
even in the presence of interactions. 
Interactions give no qualitative change to our results since 
they do not alter the relevant symmetries.  For large 
${\bm\Gamma}_{L/R}$, charging effects are smeared out in any case.
For small ${\bm\Gamma}_{L/R}$ and very strong interactions,
the result basically equals the noninteracting one (up
to a sign reversal of $I_a$ for an occupied dot).

\begin{figure}
\scalebox{0.4}{\includegraphics{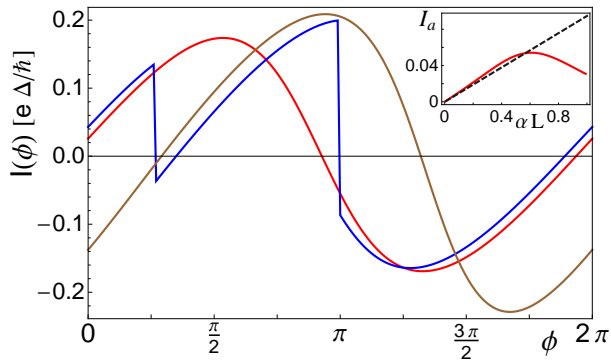}}
\caption{ 
(Color online) Numerical results for the CPR of a two-level dot
 with $\theta=0$ for $b_x = 0.3 \Delta$, $0.7 \Delta$ and
 $0.9 \Delta$  (red, blue, brown curves, respectively, 
with increasing $|I_a|$).  Parameters: $\alpha L = 0.4$, 
${\bm\Gamma}_{L,11}=2\Delta, {\bm\Gamma}_{L,22}=0, 
{\bm\Gamma}_{R,11}={\bm\Gamma}_{R,22}=\Delta/2,
\epsilon_1=-\epsilon_2=0.65\Delta$, $L=25$ nm, $\Delta=1$ meV,
and $m=0.035 m_e$.
Inset: $I_a$ from numerics (red solid curve) and 
from Eq.~(\ref{final}) (black dashed curve) vs $\alpha L$, 
with $b_x=0.5 \Delta$ and same other parameters as in main panel.
} 
\label{fig:CPR}
\end{figure}

\textit{Numerical analysis for $N=2$.---} 
We have numerically computed the full current (\ref{current}) for a 
two-level dot, taking the wave functions for a 
harmonic transverse and hard-wall longitudinal confinement
as in Ref.~\cite{luca}.  Denoting the distance between the tunnel
contacts by $L$, 
the Rashba coupling is kept fixed at a moderately small value, 
$\alpha L=0.4$, while the remaining parameters were taken  
both inside and outside the parameter regime
where the restrictions needed to derive Eq.~(\ref{final}) apply.
The resulting CPRs are shown in the main panel of Fig.~\ref{fig:CPR},
where the supercurrent is plotted for increasing values of the 
Zeeman field. These plots
show that a large value for the anomalous current $I_a$ can be
 obtained with reasonable parameters. The
hybridization matrices have been chosen to optimize the 
commutator in Eq.~(\ref{condition})
while satisfying 
${\bm\Gamma}_{L/R,12}=\sqrt{{\bm\Gamma}_{L/R,11}{\bm\Gamma}_{L/R,22}}$.
The CPR is either continuous or exhibits jumps
associated with a change of the ground state, i.e., 
with the different occupation of the relevant Andreev levels.
Such jumps are a common feature in the presence of 
Zeeman fields already without SOI \cite{jj}.
In the CPRs obtained by numerics from Eq.~(\ref{current}),
we find that the maximal (positive) current can differ from the 
minimal (negative) current in magnitude. The existence of such 
two critical currents and its potential for rectification behavior 
have been discussed in Ref.~\cite{feinberg}. 
The inset of Fig.~\ref{fig:CPR} shows a comparison between the 
full numerical results for $I_a$ and 
the analytical predictions deduced from Eq.~(\ref{final})  as a 
function of $\alpha$. The comparison is done
for an intermediate value of the Zeeman field, $b_x = 0.5 \Delta$, and
the agreement is good even for rather large $I_a$.

\textit{Conclusions.---}
We have shown that a generic model for
superconducting transport through a quantum dot exhibits spontaneously
broken TRS, leading to an anomalous supercurrent appearing at zero
phase difference between the superconductors.  The effect occurs in the
presence of spin-orbit interaction and a suitably oriented Zeeman field,
provided that the left and right contact hybridization matrices do 
not commute. This implies that at least two dot levels must 
be involved. The wide availability of mesoscopic systems holds
the promise to experimentally observe this remarkable effect 
in the near future. 

This work was supported by the SFB TR 12 of the DFG, by the
EU networks INSTANS and HYSWITCH, and by ANR-PNANO Contract MolSpintronics
No. ANR-06-NANO-27.


\begin{thebibliography}{99}
\bibitem{footnote3} 
The prediction  $I_a\neq 0$ in Ref. \cite{buzdin} is an artefact of 
an approximation violating a gauge invariance of the system.
This problem can be traced to an inconsistent use of
the quasiclassical approximation. For a correct treatment, one 
should proceed along the lines of Ref.~\cite{bezuglyi}.
\bibitem{footgauge}
The difference to Eq.~(\ref{adef}) arises because a gauge transformation
is necessary to obtain the exact solution \cite{luca}.
\end{thebibliography}
\end{document}